\documentclass[conference]{IEEEtran}
\IEEEoverridecommandlockouts
\usepackage{cite}
\usepackage{amsmath,amssymb,amsfonts}
\usepackage{algorithmic}
\usepackage{graphicx}
\usepackage{textcomp}
\usepackage{xcolor}
\def\BibTeX{{\rm B\kern-.05em{\sc i\kern-.025em b}\kern-.08em
    T\kern-.1667em\lower.7ex\hbox{E}\kern-.125emX}}
    
\usepackage{array}
\usepackage{mdwmath}
\usepackage{mdwtab}
\usepackage{eqparbox}
\usepackage{url}
\usepackage{multirow}
\usepackage{algorithmic}
\usepackage{algorithm}
\usepackage{pgfplots}
\usepackage{amssymb}
\usepackage{graphicx}
\usepackage{subfig}
\usepackage{lipsum}
\usepackage[font={small,it}]{caption}
\usepackage[font={small,it}]{caption}

\usepackage{xparse,xcolor}

\ExplSyntaxOn

\NewDocumentCommand{\setupbibcolors}{m}
 {
  \cs_set_protected:Npn \bibitem ##1
   {
    \color{ \str_case:nnF { ##1 } { #1 } { black } }
    \heba_bibitem:n { ##1 }
   }
 }

\cs_set_eq:NN \heba_bibitem:n \bibitem

\ExplSyntaxOff

\setupbibcolors{
  {baza1}{white}
  {baza2}{white}
  {baza3}{white}
   {baza4}{white}
  {baza5}{white}
   {baza6}{white}
  {baza7}{white}
   {baza8}{white}
  {baza9}{white}
   {baza10}{white}
  {baza11}{white}
  {baza12}{white}
   {baza13}{white}
  {baza14}{white}
   {baza15}{white}
  {baza16}{white}
  {baza17}{white}
  {baza18}{white}
  {Ramy1}{white}
  {Ramy2}{white}
  {Ramy3}{white}
  {Ramy4}{white}
  {Ramy5}{white}
  {Ramy6}{white}
  {Ramy7}{white}
  {Ramy9}{white}
  {Ramy10}{white}
}

\begin{document}

\font\myfont=cmr12 at 17.5pt
\title{{\myfont SoWaF: \underline{S}huffling \underline{o}f \underline{W}eights \underline{a}nd \underline{F}eature Maps: A Novel Hardware Intrinsic Attack (HIA) on Convolutional Neural Network (CNN)}\vspace{0cm}}

\author{
    \IEEEauthorblockN{
    Tolulope A. Odetola\IEEEauthorrefmark{1}, Syed Rafay Hasan\IEEEauthorrefmark{1},
    }
    %\vspace{0mm}
    \IEEEauthorblockA{
    \IEEEauthorrefmark{1}Department of Electrical and Computer Engineering, Tennessee Technological University, Cookeville, TN 38505, USA \\
    }

\vspace{-4cm}

%\and
%\IEEEauthorblockN{3\textsuperscript{rd} Given Name Surname}
%\IEEEauthorblockA{\textit{dept. name of organization (of Aff.)} \\
%\textit{name of organization (of Aff.)}\\
%%City, Country \\
%email address or ORCID}
}

\maketitle

\begin{abstract}
Security of inference phase deployment of Convolutional neural network (CNN) into resource constrained embedded systems (e.g. low end FPGAs) is a growing research area. Using secure practices, third party FPGA designers can be provided with no knowledge of initial and final classification layers. In this work, we demonstrate that hardware intrinsic attack (HIA) in such a ``secure" design is still possible. Proposed HIA is inserted inside mathematical operations of individual layers of CNN, which propagates erroneous operations in all the subsequent CNN layers that leads to  misclassification. The attack is non-periodic and completely random, hence it becomes difficult to detect. Five different attack scenarios with respect to each CNN layer are designed and evaluated based on the overhead resources and the rate of triggering in comparison to the original implementation. Our results for two CNN architectures show that in all the attack scenarios, additional latency is negligible ($<0.61\%$), increment in DSP, LUT, FF is also less than $2.36\%$. Three attack scenarios does not require any additional BRAM resources, while in two scenarios BRAM increases, which compensates with the corresponding decrease in FF and LUTs. To the authors' best knowledge this work is the first to address the hardware intrinsic CNN attack with attacker does not have knowledge of the full CNN. 
\end{abstract}

% no keywords
\begin{IEEEkeywords}
\textbf{Convolutional Neural Network, FPGA, Trojan}
\end{IEEEkeywords}

% For peer review papers, you can put extra information on the cover
% page as needed:
% \ifCLASSOPTIONpeerreview
% \begin{center} \bfseries EDICS Category: 3-BBND \end{center}
% \fi
%
% For peerreview papers, this IEEEtran command inserts a page break and
% creates the second title. It will be ignored for other modes.
\IEEEpeerreviewmaketitle

\section{Introduction}
\footnote{A version of this work will be published in ISCAS 2021}
FPGA based Convolutional Neural Network (CNN) inference has gained attention in recent times \cite{abdelouahab2018accelerating}. FPGA hardware accelerators offer good performance, high energy efficiency, fast prototyping, and capability of reconfiguration, \cite{odetola20192l}, \cite{hailesellasie2019mulnet}. The re-configurable nature of FPGAs permits flexibility in the mapping of CNNs on FPGA  with high accuracy and low precision \cite{kim2017fpga}. The adoption of High Level Synthesis (HLS) in the mapping of CNN on FPGA allows software specifications of accelerators to be synthesizable to hardware \cite{kim2017fpga}. To achieve short time-to-market, the mapping of pre-trained CNN on hardware accelerators is often outsourced to untrusted third parties. They contribute to FPGA design flow, by providing soft IPs or hard IPs (such as bitstream file). Due to their untrusted nature  hardware intrinsic security can be compromised via malicious hardware insertions, which are very difficult to detect, especially if the  IP is provided as a bitstream file. 

Different techniques of inserting hardware attacks into CNNs have been explored. Clements et. al \cite{clements2018hardware} presents a hardware Trojan framework introduced in IP designs. This hardware Trojan generates small bounded perturbations that are added to feature maps of targeted layers of the CNN and causes deterioration in the performance of CNN. Liu et. al in \cite{liu2017trojaning} discusses an attack on neural networks where samples of the input data are generated from the pre-trained model to design a trigger that activates a payload to cause misclassification. These attacks require a manipulation of the CNN parameters to achieve misclassification, which can be detected by carrying out a model integrity test on the hardware design. Moreover, traditionally it is assumed that the attacker has full knowledge of the CNN architecture. We argue that in an effort to deter hardware attacks, the project owner may hide the details of dataset by not providing details of first layer and eliminate the last layers to conceal the classification information as is the case for edge offloading for CNNs \cite{hadidi2018musical}, \cite{zeng2020coedge}.  This results in third party IP designers (potential attacker) having no means to evaluate the effectiveness and stealthiness of the attack. Hence, in this paper we demonstrate a framework of  attacks called SoWaF (Shuffling of Weights and Feature Maps to corrupt the mathematical computation) that leads to misclassification, without any knowledge of dataset and final classification layer.

\subsection{Motivational Analysis}
The above discussed attack scenario lead to the question that, what is the feasibility of hardware intrinsic attack (HIA) if  intermittent changes are made to mathematical operations of one of the layers in CNN? The premise is that if a subtle (and stealthy) minimal change in some mathematical operations can lead to misclassification, then it is extremely difficult to detect such attacks. To understand the effect of minimal change in mathematical operations we took a $3 \times 12 \times 12$ input feature map and perform convolution with a channel $3 \times 3 \times 5 \times 5$ weight matrix to obtain a $3 \times 8 \times 8$ output feature maps $O_1$. This serve as a baseline result. To device a possible attack, the channels of the weight matrix are then randomly shuffled and used to perform convolution with the input feature map to obtain another $3 \times 8 \times 8$ output feature maps $O_2$. Element wise comparison of $O_1$ and $O_2$, shows that 72\% of the values are changed more than 95\%.  This toy example inspired us to do further investigation and see the effect of SoWaF in complete CNN architecture.  
\vspace{-1mm}
\subsection{Research Challenges}
We formulated the following research challenges based on the motivational analysis:
\begin{itemize}
    \item How can an attack be triggered randomly, with the payload activated only intermittently, so that it cannot be detected easily?
    \item How the malicious changes in the mathematical operations  are implemented such that it requires minimal resources but still capable of inducing an effective attack? 
\end{itemize}

\vspace{-1mm}
\subsection{Novel Contribution and Concept Overview}
To address the aforementioned research challenges, we propose a HIA methodology for FPGA based CNN inference called SoWaF. The HIA comprises of two stages namely: Offline pre-processing and Runtime payload analysis. The attack is designed to be intermittently triggered. Overview of SoWaF methodology flow is shown in Fig. \ref{novel_diagram}. The red shaded portion of Fig. \ref{novel_diagram} shows our contribution. Section $1$ of the methodology flow involves the offline analysis of the output feature maps to design a stealthy trigger. Section $2$ shows the comparison of the additional hardware overhead incured by the HIA circuitry with the design constraints. Section $3$ shows the evaluation of the stealthiness and effectiveness of the attack. Our methodology employs the following analysis and methods:
\begin{figure}[H]
\centering
\setlength{\abovecaptionskip}{0mm}   % 0.5cm as an example
\setlength{\belowcaptionskip}{0mm}   % 0.5cm as an example
\centering
\includegraphics[height=0.28\textwidth]{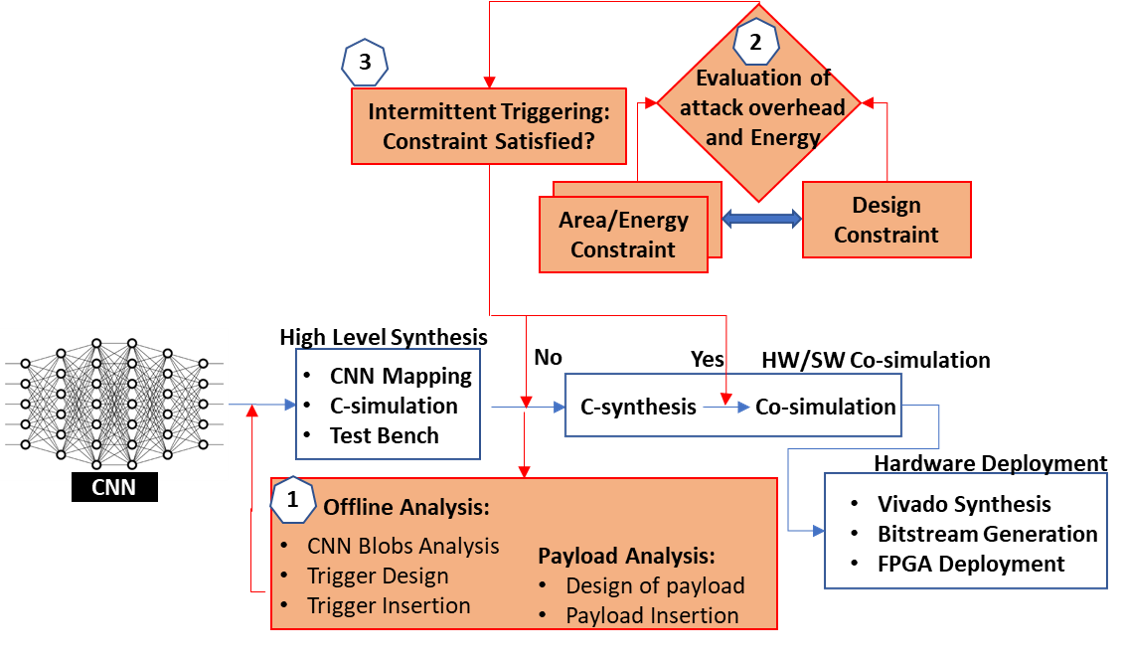}
\caption{Design time flow of the HIA methodology. Highlighted boxes represent the novel contributions}
\vspace{0mm}
\label{novel_diagram}
\end{figure}
\vspace{-1em}
\begin{itemize}
    \item To design a trigger, we propose the exploitation of computation of the layer-by-layer feature maps.
    \item To achieve stealthiness, we propose a positional relationship based probabilistic trigger, which is intermittent. 
    \item To achieve misclassification with minimum resource overhead, we propose a novel payload that disrupts certain mathematical operations with very minimal (if any) added resources.
\end{itemize}
The remainder of this paper is organized as follows: Section II describes the threat model. Section III discusses the proposed attack design. Section IV  shows experimental results and discussion. Section V provides comparison with state-of-the-art and Section VI concludes the paper.
\vspace{-2mm}

\section{Threat model}
\begin{figure}[]
\centering
\setlength{\abovecaptionskip}{0mm}   % 0.5cm as an example
\setlength{\belowcaptionskip}{0mm}   % 0`.5cm as an example
\centering
\includegraphics[height=0.21\textwidth]{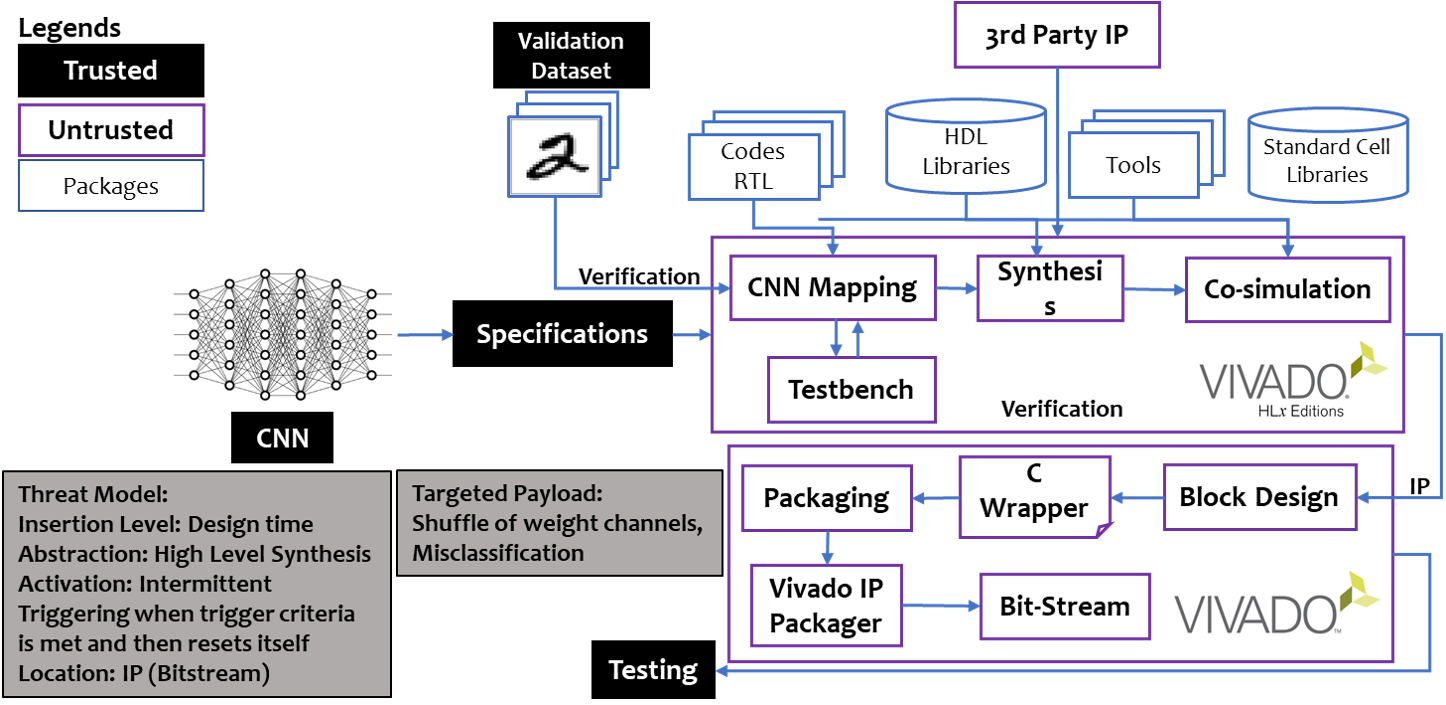}
\caption{Brief overview of the mapping of CNN to FPGA elaborating on the threat model and the corresponding payloads.}
\vspace{0mm}
\label{threat_model}
\end{figure}

This work proposes a gray-box attack where the attacker has little knowledge of the CNN architecture. We assume that the third party IP designer is not trustworthy. It is assumed that the attacker has no access of the training and testing data samples, i.e. attacker is only designing a CNN without its head (last layers) and initial layers. The attacker provides the implemented CNN hardware design as a bitstream file to the defender (project owner).
Fig. \ref{threat_model} shows the trusted and untrusted sections of the design. The 3rd party designer has access to the untrusted sections based on specifications and requirements provided by the trusted party. It is also assumed that for verification purposes, a hardware validation dataset is provided to the 3rd party designer (a normal industrial practice \cite{odetola20192l}), without revealing any information about the initial layers.   
\vspace{-5mm}

\section{SoWaF Attack Methodology}
 The proposed methodology is sub-divided into 2 phases namely:
  \begin{itemize}
      \item Offline Pre-processing: Trigger Design
      \item Runtime: Payload Operation
  \end{itemize}

\subsection{Offline Pre-processing: Trigger Design}
\vspace{-4mm}
\begin{figure}[h]
\centering
\setlength{\abovecaptionskip}{0mm}   % 0.5cm as an example
\setlength{\belowcaptionskip}{0mm}   % 0`.5cm as an example
\includegraphics[height=0.265\textwidth]{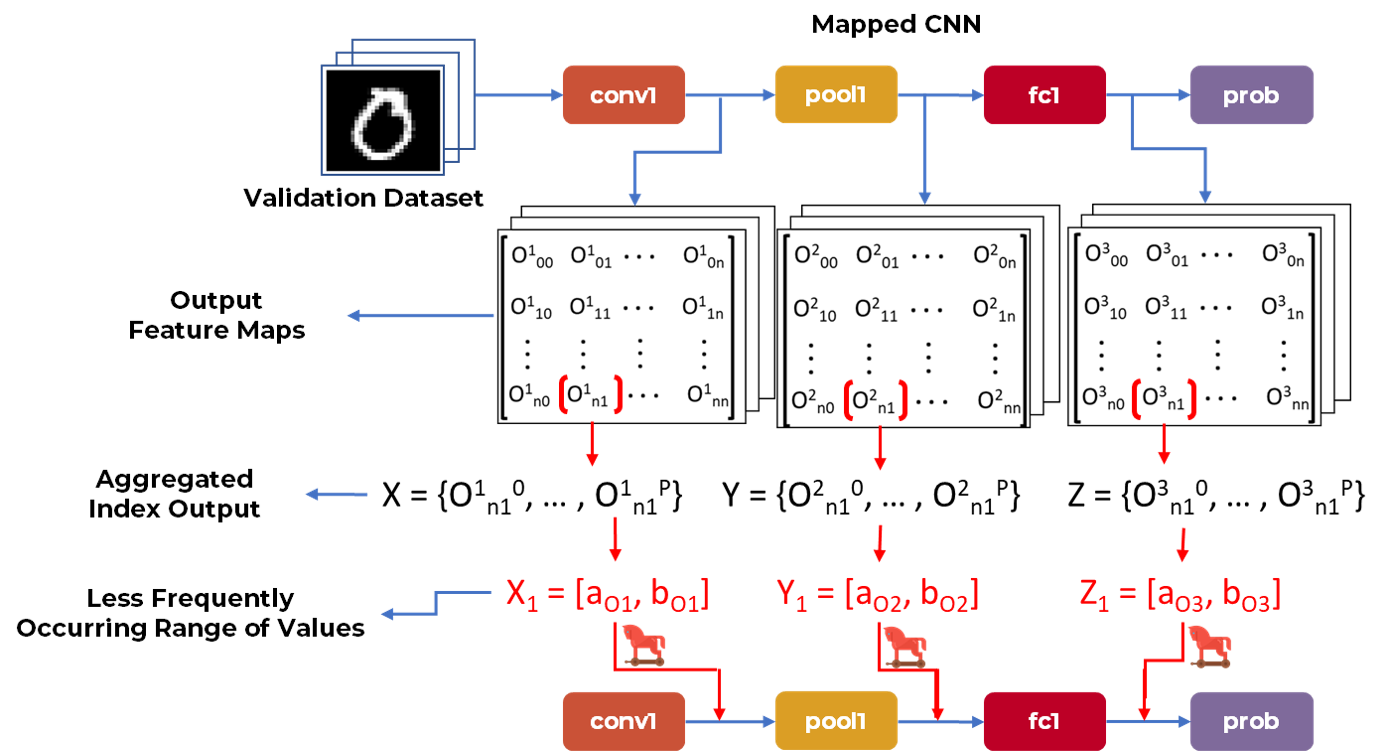}
\vspace{0mm}

\caption{Offline pre-processing (Trigger Design): Conceptual Representation of offline processing: the aggregation of values from a chosen index helps the selection of less occurring range of values (RoV) (triggers) and the insertion of the trigger (where $P$ is the size of the Validation dataset, $O^1$, $O^2$, $O^3$ are the output feature maps of $conv1$, $pool1$ and $fc1$ layers, respectively.) }
\vspace{0mm}
\label{offline}
\end{figure}

\vspace{-2mm}
During the functional verification stage the hardware validation dataset can be used by the attacker to access the respective CNN layer's output feature maps for all the dataset. As illustrated in Fig. \ref{offline}, the attacker assess the statistical properties of the output feature maps to setup a trigger.

In this work, to design the trigger we randomly choose an index ($O^w_{n,m}$, where $w$ is the layer, tuple ($n,m$) represents the rows and columns of the index) of one of the randomly chosen channels  of the output feature map of a targeted layer as shown in line 1 of Algorithm \ref{alg:MYALG}. During verification, the attacker may monitor the the values ($X$ or $Y$ or $Z$) as shown in Fig. \ref{offline} of the randomly selected index ($O^w_{n,m}$) against the validation dataset to obtain the range of values (RoV) (where $[a_w, b_w]$ represent the minimum and maximum value of RoV respectively) that are likely to occur at a particular index. This serves as a sample space to estimate RoV that occur less often on the chosen index as shown in line 3 - 5 of Algorithm \ref{alg:MYALG}. From the Aggregated index outputs ($\{O^{w^0}_{n,m}, ... O^{w^P}_{n,m}\}$, where $0, ... P$ represent each data instance in the validation dataset), as shown in Fig. \ref{offline}, of a chosen index, we select a RoV whose number of occurrence in the validation dataset satisfy a chosen threshold $(T(O^w_{n,m}) \rightarrow c([a_w, b_w]) = M)$  as shown in Line 6 - 10 of Algorithm \ref{alg:MYALG}. The selected RoV ($[a_w, b_w]$) for a given CNN layer serve as the trigger for the HIA. This offline pre-processing algorithm, Algorithm \ref{alg:MYALG}, enables the proposed attack to assess the RoV and to select a stealthy trigger while processing an image. 
%________________________ Algorithm ________________________%
\begin{algorithm}[H]
\caption{Offline Pre-processing: Trigger Design}
\begin{algorithmic} [1]
\REQUIRE Mapping of the CNN to the desired hardware in HLS (C++).
\REQUIRE Verification of mapped CNN hardware design.
\STATE Select CNN layer index $O^w_{n,m}$
\FOR {each image (X) $\in$ validation dataset (of size $P$)}
\STATE $A = \{O^{w^0}_{n,m}, ... O^{w^P}_{n,m}\} \in O^w_{n,m}$ \\
    where: $\{O^{w^0}_{n,m}, ... O^{w^P}_{n,m}\}$ are the numerical values of the chosen index for each data instance in the validation dataset ($1$, $2$, ... $P$)\\
    $A$ is output feature map of any layer ($X$ or $Y$ or $Z$) \\
    $O^w$ is the chosen channel of the targeted layer\\
    $n,m$ are the row/column indexes of the chosen channel\\
    $w$ represents the targeted layer\\
\ENDFOR
\STATE Select less frequently occurring RoV $[a_w, b_w]$ from $A$
\IF {$O^w_{n,m}$ : T($O^w_{n,m}$)}
\STATE Select $[a_w, b_w]$ from $A$ \\
     where: $T(O^w_{n,m}) \rightarrow c([a_{w}, b_{w}) = M$\\
    $c$ = number of elements in $A$ within $[a_w, b_w]$\\
    $M$ = Chosen threshold for $c$\\
    $T$ is the function that returns a Boolean if the numerical value of the index satisfies the chosen RoV\\
    $[a_w, b_w]$ are the lower/upper limit of the chosen RoV\\
\ELSE
\STATE Select new range $[a_w, b_w]$ and repeat steps 4 to 6.
\ENDIF
\STATE Insert $[a_w, b_w]$ as trigger in mapped CNN
\end{algorithmic}
\label{alg:MYALG}
\end{algorithm}
%_____________________________________________________________%
\vspace{-4mm}

\subsection{Runtime: Payload Operation}
\begin{figure}[t]
\centering
\setlength{\abovecaptionskip}{0mm}   % 0.5cm as an example
\setlength{\belowcaptionskip}{0mm}   % 0`.5cm as an example
\includegraphics[height=0.35\textwidth]{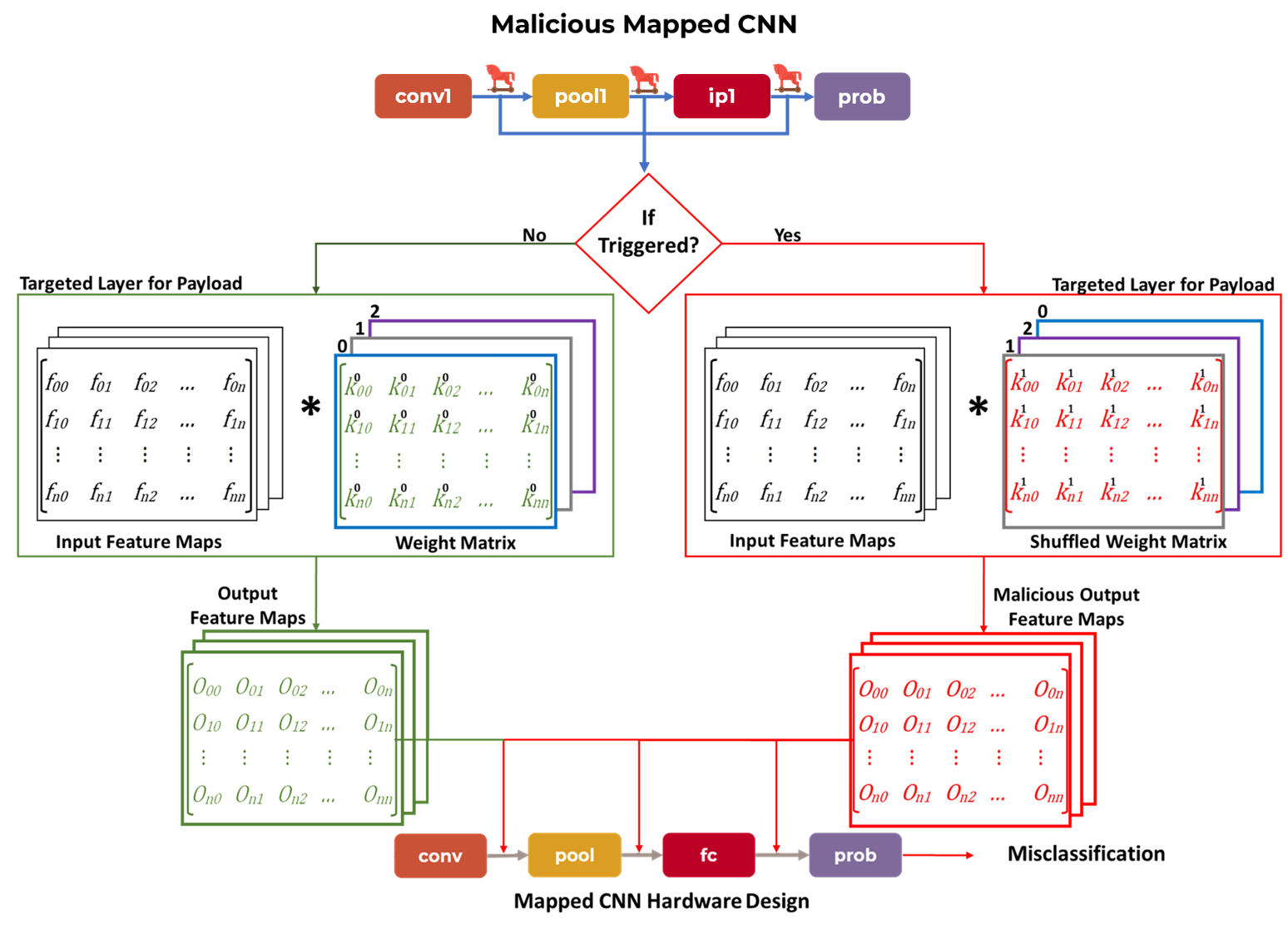}
\caption{Representation of the payload during runtime: the right side of the image (shaded in red) shows the weight matrix channels reshuffled to obtain malicious feature maps to achieve misclassification}
\label{runtime}
\end{figure}

To design the payload we proposed an algorithm,  Algorithm \ref{alg:MYALG3}. The payload monitors the selected CNN layer, the selection of CNN layer depends on the additional resource overhead incurred due to the targeted layer and the rate of triggering on the layer. Upon triggering, for convolution layers, the payload shuffles the channels of the weight matrix with another one as illustrated on the right hand side of the decision block in Fig. \ref{runtime}. Line 4 of Alogrithm \ref{alg:MYALG3} makes sure that the individual channels in a particular layer are swapped, i.e. $Q_w[j_0]$ is swapped with $Q_w[j_f]$ (where $0 < f < l$ and $l$ is the amount of weight matrix channels). Because CNN layers other than convolution and fully connected layers (such as Pooling, etc.) do not have weight matrices and channels, therefore the are treated separately in line 5 of Algorithm \ref{alg:MYALG3}. This payload is empirical crafted  offline with several experiments (explained in Section IV) to make sure that the malicious mathematical modifications likely results in misclassification.

%________________________ Algorithm ________________________%
\begin{algorithm}
\caption{Runtime: Payload Operation}
\begin{algorithmic} [1]
\REQUIRE CNN Deployment.
\FOR {each image cycle (Im)}
\STATE Monitor selected CNN layer index $O^w_{n,m}$
\IF {$O^w_{n,m} \in [a_w, b_w]$ }
\STATE $Q_w[j_0, j_1, ... j_l] =  Q_w[j_f, j_{f+1}, ... j_ {l-f}]$ (conv layer) 
\STATE $R_w[j_0, j_1, ... j_l] =  R_w[j_f, j_{f+1}, ... j_{l-f}]$ (pool layer)
   \\where: $f$ is the order factor of shuffling
   \\$Q_w[j_0, j_1, ... j_f]$ is the  default weight matrix order
   \\ $R_w[j_0, j_1, ... j_f]$ is the  default output channel order 
   \\ $l$ is the number of weight channels or output channels
\IF {$0$ $< j <$ $\frac{l}{2}$}
    \STATE $\frac{l}{2} <  f < l$ \\
    \ELSE
    \STATE $0 < f < \frac{l}{2}$ 
    \ENDIF
\ENDIF
\ENDFOR
\end{algorithmic}
\label{alg:MYALG3}
\end{algorithm}
\vspace{0mm}
%_____________________________________________________________%

\begin{table*}[]
\centering
\caption{Resource overhead comparison between attacks on different layers of LeNet and LeNet-3D compared to their respective originals}
\begin{tabular}{|c|c|c|c|c|c|c|c|c|c|c|c|c|}
\hline
Network                                                                               & \begin{tabular}[c]{@{}c@{}}Attack Scenario (Sn):\\ Layer\end{tabular} & Chs & BRAM & \begin{tabular}[c]{@{}c@{}}\% \\ diff\end{tabular} & DSPs & \begin{tabular}[c]{@{}c@{}}\% \\ diff\end{tabular} & \begin{tabular}[c]{@{}c@{}}LUTs \\ (x1000)\end{tabular} & \begin{tabular}[c]{@{}c@{}}\% \\ diff\end{tabular} & \begin{tabular}[c]{@{}c@{}}FFs \\ (x1000)\end{tabular} & \begin{tabular}[c]{@{}c@{}}\% \\ diff\end{tabular} & \begin{tabular}[c]{@{}c@{}}Latency (x1000)\\ clock-cycles\end{tabular} & \begin{tabular}[c]{@{}c@{}}\% \\diff\end{tabular} \\ \hline
\multirow{6}{*}{LeNet}                                                                & Original                                                              & -   & 42   & -                                                  & 33   & 0                                                  & 118.5                                                   & -                                                  & 58.3                                                   & -                                                  & 680.4                                                               & -                                                        \\ \cline{2-13} 
                                                                                      & Sn1: $conv1$ attack                                                   & 6   & 42   & 0                                                  & 33   & 0                                                  & 119.2                                                   & +0.61                                              & 59.2                                                   & +1.5                                               & 680.51                                                              & +0.003                                                       \\ \cline{2-13} 
                                                                                      & Sn2: $pool1$ attack                                                   & 6   & 42   & 0                                                  & 33   & 0                                                  & 118.9                                                   & +0.34                                              & 58.8                                                   & +0.76                                              & 680.51                                                              & +0.003                                                       \\ \cline{2-13} 
                                                                                      & Sn3: $conv2$ attack                                                   & 16  & 53   & +26                                                & 33   & 0                                                  & 121.3                                                   & +2.36                                              & 58.8                                                   & +0.81                                              & 680.58                                                              & +0.013                                                      \\ \cline{2-13} 
                                                                                      & Sn4: $pool2$ attack                                                   & 16  & 42   & 0                                                  & 33   & 0                                                  & 119.2                                                   & +0.34                                              & 59.3                                                   & +0.76                                              & 680.51                                                              & +0.003                                                       \\ \cline{2-13} 
                                                                                      & Sn5: $conv3$ attack                                                   & 120 & 162  & +285                                               & 33   & 0                                                  & 780.7                                                   & -34                                                & 34.5                                                   & -41                                                & 680.74                                                              & +0.038                                                      \\ \hline\hline
\multirow{6}{*}{\begin{tabular}[c]{@{}c@{}}LeNet-3D\\ for  \\ Cifar10\end{tabular}} & Original                                                              & -   & 59   & -                                                  & 37   & -                                                  & 49.0                                                    & -                                                  & 39.7                                                   & -                                                  & 1685.71                                                             & -                                                        \\ \cline{2-13} 
                                                                                      & Sn1: $conv1$ attack                                                   & 5   & 59   & 0                                                  & 37   & 0                                                  & 49.9                                                    & +1.81                                              & 40.5                                                   & +1.8                                               & 1685.73                                                             & +0.001                                                     \\ \cline{2-13} 
                                                                                      & Sn2: $pool1$ attack                                                   & 5   & 59   & 0                                                  & 37   & 0                                                  & 49.6                                                    & +1.16                                              & 40.4                                                   & +1.76                                              & 1685.72                                                             & +0.001                                                    \\ \cline{2-13} 
                                                                                      & Sn3: \$conv2 attack                                                   & 20  & 79   & +34                                                & 37   & 0                                                  & 48.6                                                    & -0.78                                              & 39.0                                                   & -1.9                                               & 1685.72                                                             & +0.001                                                    \\ \cline{2-13} 
                                                                                      & Sn4: $pool2$ attack                                                   & 20  & 59   & 0                                                  & 37   & 0                                                  & 50.0                                                    & +1.93                                              & 41.0                                                   & +3.2                                               & 1695.99                                                             & +0.61                                                      \\ \cline{2-13} 
                                                                                      & Sn5: $conv3$ attack                                                   & 100 & 159  & +169                                               & 37   & 0                                                  & 20.1                                                    & -59                                                & 10.0                                                   & -74.6                                              & 1685.72                                                             & +0.001                                                    \\ \hline
\end{tabular}
\label{model_results}
\end{table*}
\vspace{-1mm}

\begin{figure}[h]
\centering
\setlength{\abovecaptionskip}{0mm}   % 0.5cm as an example
\setlength{\belowcaptionskip}{0mm}   % 0`.5cm as an example
\includegraphics[width=0.49\textwidth]{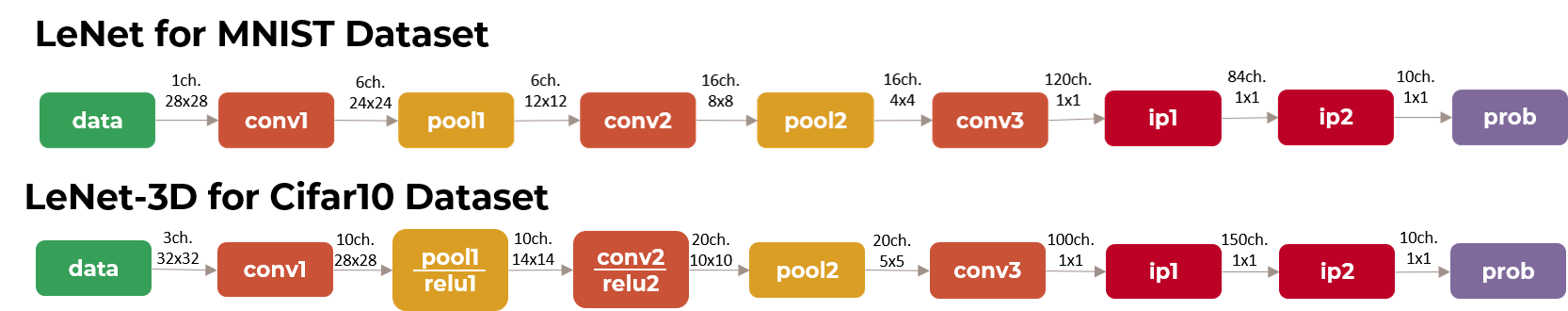}
\vspace{0mm}
\caption{LeNet and LeNet-3D CNN Models}
\vspace{0mm}
\label{models}
\end{figure}
\vspace{-5mm}

\section{Experiment Setup, Results and Discussion}
 The mapped CNN IP is designed using Xilinx's Vivado and Vivado HLS 2018.3 and to generate an IP for resource constrained devices. Vivado is used to integrate  the  generated IP with  AXI-interconnects  and  ZYNQ processor (FPGA ZCU 7020 with clock frequency 100MHz). The HIA is implemented on Lenet (Fig. \ref{models}) trained on MNIST dataset and LeNet-3D for Cifar10 datasets, respectively. In this work, we propose $5$ different scenarios, where each layer  (from $conv1$ to $conv3$) is infected with the HIA. Stealthiness (defined as of the additional hardware  overhead  (BRAM,  DSP,  flip-flops (FFs),  look-up tables (LUTs), Latency) is evaluated for each case  and effectiveness (defined as the rate of triggering) of the inserted HIA compared to the original implementation of the mapped CNNs. The attack is carried out on the convolution and pooling layers.
 
 \begin{figure}[]
\centering
\setlength{\abovecaptionskip}{0mm}   % 0.5cm as an example
\setlength{\belowcaptionskip}{0mm}   % 0`.5cm as an example
\centering
\includegraphics[height=0.21\textwidth]{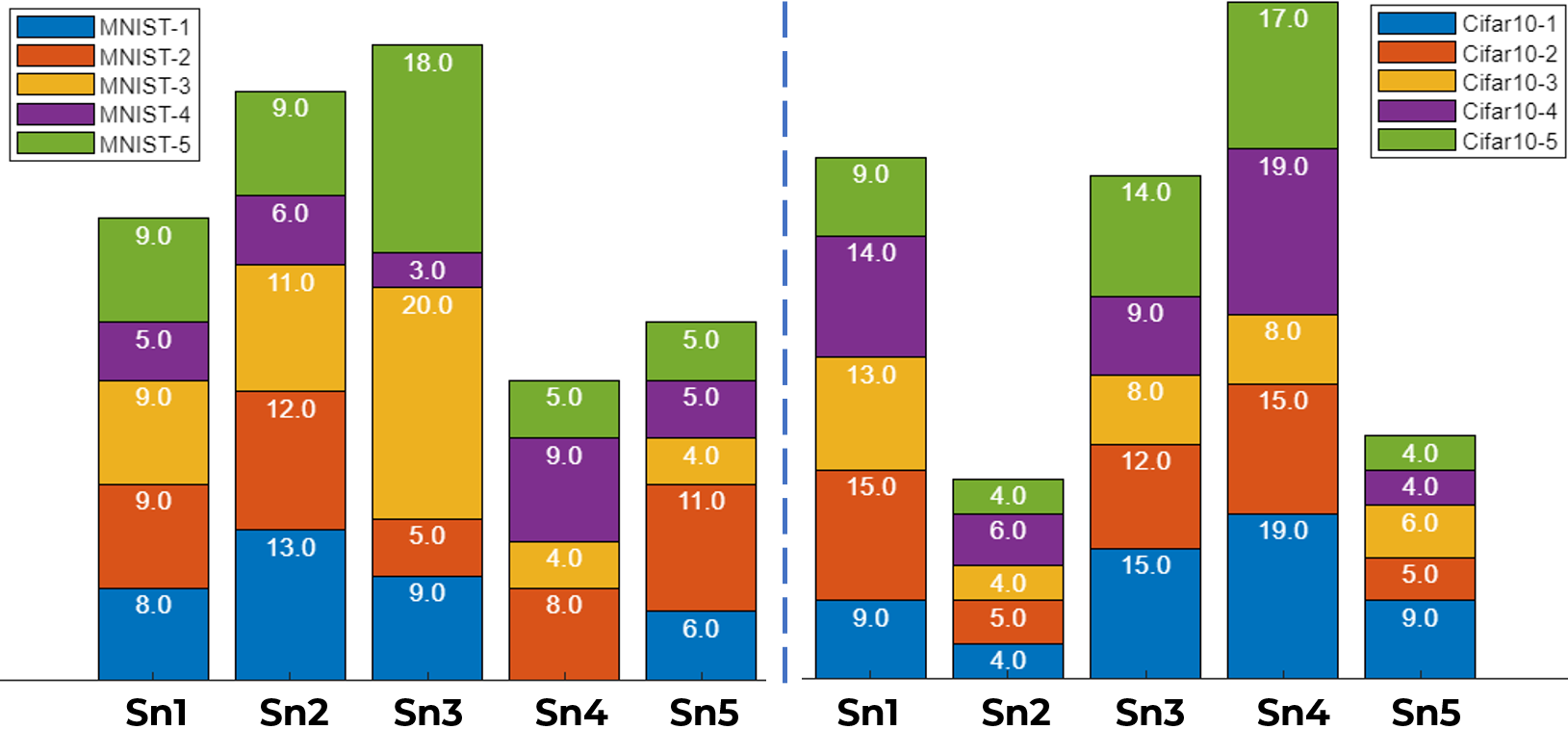}
\caption{Graph showing the random nature and low-triggering rate of the different attack scenarios: \textbf{Left} represents LeNet trained on MNIST and \textbf{Right} and LeNet-3D CNN model}
\label{stealth}
\end{figure}

Table \ref{model_results} shows that DSP remains same for all the attack scenarios. In both CNNs, BRAM usage remains the same except for $Sn3$ and $Sn5$, where BRAM are increased ($5th$ column in Table \ref{model_results}). $Sn5$ compensates this with lower LUTs and FFs usage (see columns $9$ and $11$ in Table \ref{model_results}). For LUTs and FFs in all the scenarios, other than $Sn5$, (i.e. $Sn1$ - $Sn4$) have a very modest increment in usage (up to $2.36\%$). Similarly, difference in latency between designs with and without HIA design is less than $0.61\%$ in $Sn1$ - $Sn5$ (last column in Table I). Hence, we conclude that $Sn1$,$Sn2$ and $Sn4$ can be good choices for an attacker for a stealthy attack, as overall resources and latency effects is minimal. It is observed that for both CNN models, number of weight and output feature map channels are proportional to the additional amount of hardware resources overhead. This can be confirmed from the results of $Sn3$ and $Sn5$ where the higher number of output feature map channels has resulted in higher memory usage. To demonstrate the randomness of the proposed attack, various random input validation dataset is examined. In Fig. \ref{stealth}, for the $Sn1$, when five sets (200 images each) of data is provided to LeNet and LeNet-3D, the number of trigger occurrences vary randomly between 5 to 9. Same is true for other attack scenarios- making our attack random and stealthy. 

%\vspace{-3mm}

%\vspace{-2mm}
\section{Comparison with State-of-the-Art}
%\vspace{-3mm}
 %Other approaches have been adopted to perform hardware Trojan insertion in pre-trained CNN implemented on hardware accelerators. Zou et al. \cite{zou2018potrojan} proposes PoTrojan, a hardware Trojan on pre-trained models. Clements et. al \cite{clements2018hardware} proposes an attack on neural networks that is targeted at specific layer(s). In this case, adversarial images to fool the CNN are generated and used to determine the required perturbations to be added to the output feature maps to cause mis-classification. Clements et. al \cite{clements2019hardware} proposes an attack on neural networks that is targets computational blocks or activation function of specific CNN layer(s) to cause mis-classification.  Zhao et. al \cite{zhao2019memory} proposes a memory Trojan attack on hardware accelerators. Liu et. al \cite{liu2020sequence} proposes a hardware Trojan that is sequence triggered based on the order in which input images are passed through the CNN. These approaches requires full knowledge of the CNN architecture to appropriate evaluate the effectiveness of the attack. 
 We summarized these differences in Table \ref{comparison1}. Most of the state-of-the-art hardware/firmware attacks on the hardware deployment of CNN requires full knowledge of CNN architecture [5], [7], [8], and [10]. In this paper we argue that to deter the hardware attackers first and last layers may be kept hidden from the un-trusted designers. Hence, our proposed attack is made under more constrained condition. In addition, existing literature requires actual input image for triggering [5], [7] - [10], while proposed design just make use of validation data set. Also in the proposed design, payload implementation does not require extra computation, unlike [5], [7], [8].

\begin{table}[h]
\centering
\caption{Comparison of our approach with other techniques}
\begin{tabular}{|c|c|c|c|c|c|c|}
\hline
Criteria                               & \cite{clements2018hardware} &  \cite{zou2018potrojan}& \cite{clements2019hardware} & \cite{zhao2019memory} & \cite{liu2020sequence} & Ours \\ \hline
Req. full CNN arch.     &        \checkmark           &                    \checkmark      &      \checkmark                    &   x                 &    \checkmark    & x\\ \hline
Req. changes in the weights &     x                &    \checkmark                      &         x             &        x          &    x  & x\\ \hline
Trig. req. Input Image              &        \checkmark             &      \checkmark                   &             \checkmark              &         \checkmark            &     \checkmark    & x \\ \hline
Payload req. extra computation               &       \checkmark             &      \checkmark                    &             \checkmark              &          x          &     x & x\\ \hline
\end{tabular}
\label{comparison1}
\end{table}
\vspace{-3mm}

\footnote{Several works have addressed security and privacy in many applications~\cite{baza1,baza2,baza3,baza5,baza4,baza6,baza7,baza8,baza9,baza10,baza11,baza12,baza13,baza14,baza15,baza16,baza17,Ramy1,Ramy2,Ramy3,Ramy4,Ramy5,Ramy6,Ramy7,Ramy9,Ramy10}}

\section{Conclusion}
To the best of authors' knowledge, this is the first work to propose  a  HIA  targeted at FPGA based CNN inference  with attacker having no knowledge of initial layers, datasets, and final classification layer. The attack achieves misclassification by shuffling the weight matrices of convolution layers to propagate wrong feature maps. This attack is carried out without changes in the model parameters. Our results for two CNN architectures show that in all the attack scenarios, additional latency is negligible ($<0.61\%$), increment in DSP, LUT, FF is also less than $2.36\%$. Three of the five investigated scenarios show very minimal changes in BRAM. Proposed attack is triggered intermittently and our results show that the number of triggers and its occurrence instance are completely random.

\ifCLASSOPTIONcaptionsoff
  \newpage
\fi

\bibliographystyle{IEEEtran}
\bibliography{references2.bib}

\end{document}